\documentclass[12pt,preprint]{aastex}
\newcommand{\msini}{\ensuremath{m \sin{i}}}
\newcommand{\feh}{\ensuremath{[\mbox{Fe}/\mbox{H}]}}
\newcommand{\rphk}{\ensuremath{\mbox{R}'_{\mbox{\scriptsize HK}}}}
\newcommand{\mv}{\ensuremath{\mbox{M}_{\mbox{\scriptsize V}}}}

\newcommand{\dmv}{\ensuremath{\Delta\mv}}
\newcommand{\teff}{\ensuremath{T_{\mbox{\scriptsize eff}}}}
\newcommand{\persec}{\ensuremath{\mbox{s}^{-1}}}

\newcommand{\mjup}{\ensuremath{\mbox{M}_{\mbox{\scriptsize Jup}}}}
\newcommand{\mearth}{\ensuremath{\mbox{M}_{\earth}}}

\def\astrosun {\mbox{$\odot$}}
\newcommand{\Msol}{\ensuremath{\mbox{M}_{\astrosun}}}

\shorttitle{A Jupiter Twin} \shortauthors{Wright et al.}

\begin{document}
\title{The Jupiter Twin HD~154345{\it b}\altaffilmark{1}}
\altaffiltext{1}{Based on observations obtained at the W. M. Keck
  Observatory, which is operated jointly by the University of
  California and the California Institute of Technology.  The Keck
  Observatory was made possible by the generous financial support of
  the W. M. Keck Foundation.}  \author{J. T. Wright\altaffilmark{2},
  G. W. Marcy\altaffilmark{3,4}, R. P. Butler\altaffilmark{5},
  S. S. Vogt\altaffilmark{6}, G. W. Henry\altaffilmark{7},
  H. Isaacson\altaffilmark{8}, and A. W. Howard\altaffilmark{3,9}}

\altaffiltext{2}{Department of Astronomy, 610 Space Sciences Building,
  Cornell University, Ithaca, NY 14853\\jtwright@astro.cornell.edu}

\altaffiltext{3}{Department of Astronomy, 601 Campbell Hall,
  University of California, Berkeley, CA 94720-3411}

\altaffiltext{4}{gmarcy@berkeley.edu} 

\altaffiltext{5}{Department of
  Terrestrial Magnetism, Carnegie Institute of Washington, 5241 Broad
  Branch Road NW, Washington, DC 20015-1305\\butler@dtm.ciw.edu}

\altaffiltext{6}{UCO/Lick Observatory, University of California, Santa
  Cruz, CA 95064\\vogt@ucolick.edu} 

\altaffiltext{7}{Center of
  Excellence in Information Systems, Tennessee State University, 3500
  John A. Merritt Blvd., Box 9501, Nashville, TN 37209,
  USA\\henry@schwab.tsuniv.edu} 

\altaffiltext{8}{Department of Physics
  and Astronomy, San Francisco State University, San Francisco, CA
  94132\\isaacson@stars.sfsu.edu} 

\altaffiltext{9}{Townes Fellow,
  Space Sciences Laboratory, UC Berkeley\\howard@astro.berkeley.edu}

\notetoeditor{It is important that the final letter in the title be
  lower case.}

\begin{abstract}
We announce the discovery of a twin of Jupiter orbiting the slightly
metal-poor ($\feh = -0.1$) nearby (d$= 18$ pc) G8 dwarf HD 154345.
This planet has a minimum mass of 0.95 \mjup\ and a 9.2 year, circular
orbit with radius 4.2 AU.  There is currently little or no evidence
for other planets in the system, but smaller or exterior planets
cannot yet be ruled out.  We also detect a $\sim$ 9-year activity
cycle in this star photometrically and in chromospheric emission.  We
rule out activity cycles as the source of the radial velocity
variations by comparison with other cycling late-G dwarfs.
\end{abstract}


\section{Introduction}

One of the primary goals of our planet search at Keck Observatory,
which began collecting precise radial velocities (RVs) of nearby
Sun-like stars in earnest in April 1997, was to discover the first
signpost of a true Solar System analog, a Jupiter twin.  This would
require high precision ($< 3$ m/s) measurements stable over $10-20$
years of regular observation to detect the 12 m/s (modulo $\sin{i}$),
sinusoidal signal induced by a 1 \mjup\ object over a $\sim 10$ year
period.  A survey would also have to measure the effects of stellar
activity cycles on the long-term RV stability of Sun-like stars, since
such cycles have periods of $\sim 10$ years, similar to those of the
orbits of Jupiter-type planets.

Now, with the recent passage of the 10 year anniversary of our first
observations at Keck, we announce that we have achieved these goals
and present the first veritable Jupiter analog.

The 10 year baseline of precise velocities has already revealed
several long-period planets, including the following with $P > 8$ yr:
an $\msini =3$ \mjup\ planet in an eccentric orbit around HD 72659
\citep{Butler03}; planets in eccentric orbits with interior gas giants
around HD 190360 and HD 217107 \citep{Vogt05}; and 55 Cnc {\it d}
\citep{Marcy_55cnc}, an $\msini = 4$ \mjup\ planet in a circular orbit
and the outermost member of an extraordinary 5-planet system (Fischer
et al. 2008, {\it ApJ} accepted).  The {\it Catalog of Nearby
  Exoplanets}\footnote{http$://$exoplanets.org/} \citep{Butler06}
counts 12 other nearby planets with $P > 6$ yr discovered to date.

Characterization of these planets is complicated by their long
periods: while normally we prefer to observe multiple orbits before
announcing a planet so that we can separate the effects of additional
planets in the system, such caution can be impractical when orbital
periods exceed 8 years.  \citet{Wright07a} developed techniques for
constraining the minimum mass of long-period planets, including HD
154345{\it b}, whose orbits were apparently nearly complete but
nonetheless lacked well-constrained periods.  Since that work's
publication, HD 154345{\it b} has completed its orbit, revealing its
minimum mass, period, and eccentricity all to be at the low end of the
range of plausible values listed in \citet{Wright07a}.

HD 154345 joins 55 Cnc in demonstrating that the architecture of the
Solar System -- a dominant, Jupiter-mass planet at 4--5 AU in a
circular orbit with only lower-massed objects interior -- while rare,
is not unique.

\section{The HD~154345 System}
HD~154345 (= GJ 651) is a bright, nearby, somewhat metal-poor G8
dwarf.  Its Mount Wilson $S$-index of 0.18 is only slightly higher
than the minimum seen for stars of similar \teff, metallicity, and
evolutionary status (J. T. Wright, in prep.), indicating that it is
likely an old (age $> 2$ Gyr), slowly-rotating \footnote{The canonical
  age-rotation period formula of \citet{Noyes84} yields a rotation
  period of 31 d, a period not seen in the periodogram of the RV
  residuals} field star. Table~\ref{star} contains compiled stellar
data for HD~154345 from the SPOCS catalog \citep{SPOCS}, {\it
  Hipparcos} \citep{PerrymanESA}, and the activity catalog of
\citet{Wright04}.  Being metal poor, HD 154345 sits just below the
main sequence, (it has negative \dmv) as computed in \citet{Wright05}.

Stars of this color and activity level typically exhibit r.m.s radial
velocity variations of 2-5.5 m \persec \citep{Wright05}, an estimate
which includes systematic errors as well as any astrophysical noise
intrinsic to the star.  This is an overestimate for the data presented
here, because of the large number of data points taken since upgrades
at HIRES in August 2004 decreased our systematic errors down to 1.5 m
\persec\, at worst.  We thus adopt a uniform ``jitter'' estimate of
2.5 m \persec\ for this work.

In Table~\ref{vels} we present the 55 radial velocity measurements we
have collected for this star since 1997, using the iodine technique
\citep{Butler96b}.  Typical exposure times in good weather were $\sim
1$ minute.

\begin{figure}[!ht]
\plotone{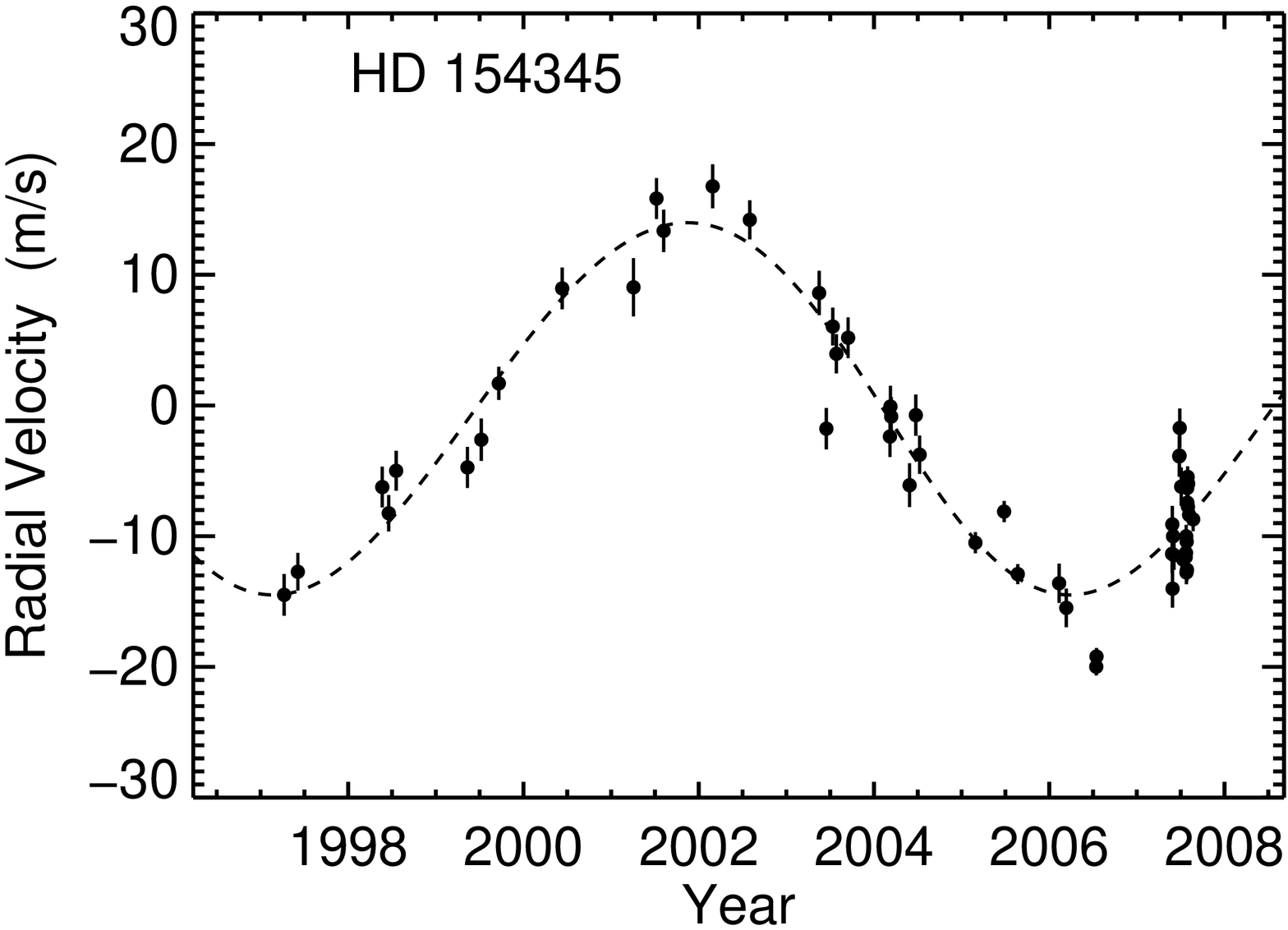}
\caption{Radial velocity vs. time for HD~154345.}\label{154345}
\end{figure}

We have fit the data under the assumption that HD 154345{\it b} is the
only planet in the system, binning the data in 3-day bins before
adding our jitter estimate in quadrature with random errors.  The
velocities are consistent with an $\msini \sim 1$ \mjup\ planet in a
circular 4.2 AU orbit and no other companions.  Lower jitter estimates
yield a best fit with slightly larger eccentricity.  Fig.~\ref{154345}
shows the best fit, and Table~\ref{orbit} contains the best-fit
parameters, with errors computed using the bootstrapping technique
described in \citet{Butler06}.

\section{Activity Cycles in Late G Dwarfs}
\begin{figure}[!ht]
\plotone{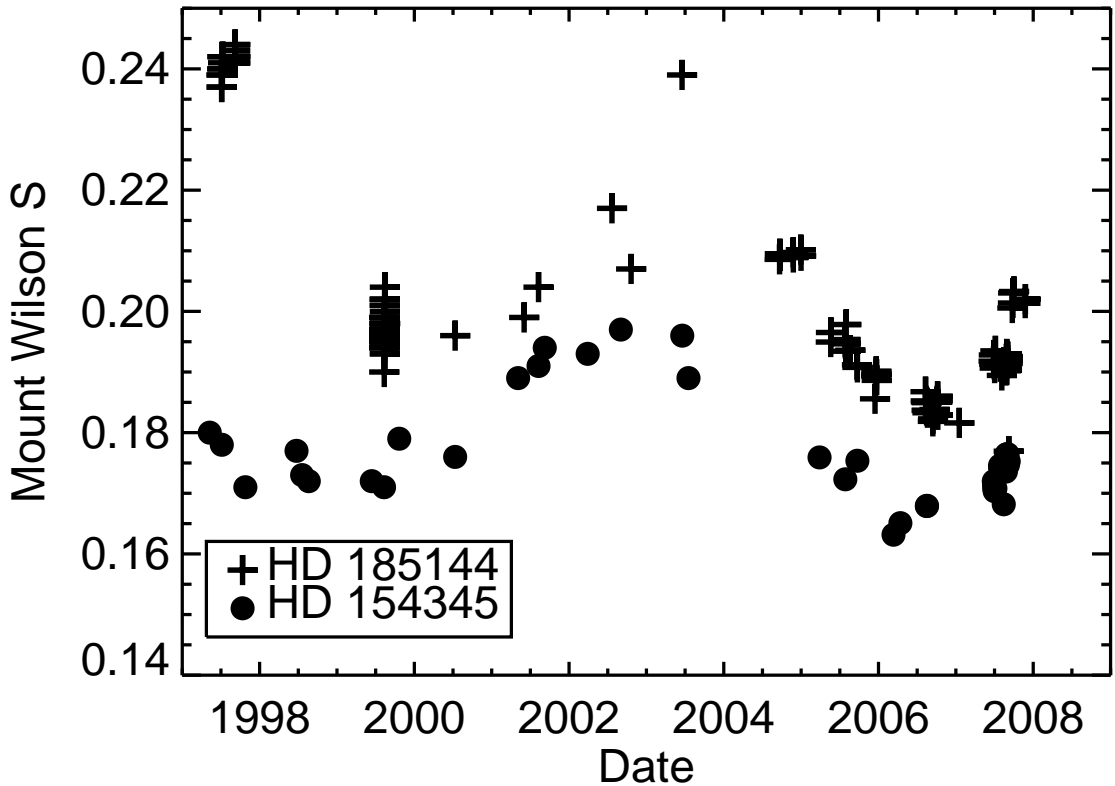}
\caption{Mount Wilson activity index measured from RV science spectra
  taken at Keck Observatory.  Except for a few small discrepancies,
  the temporal coverage of these data are the same as that of the RV
  data for both of these stars.  Data from prior to 2004 are taken
  from the ``differential'' measurements in \citet{Wright04};
  subsequent data have been extracted in a similar manner
  (H. Isaacson, in preparation).  Both HD 154345 and the RV-stable
  star HD 185144 show strong evidence of activity cycles, though the
  cycle strength and overall activity level in HD 185144 is
  considerably larger.  Cycles such as these are not uncommon in old G
  dwarfs, typically have $\sim$ 10 year periods, and are not observed
  to have an effect on long-term RV stability.  Data for the two stars
  are plotted on the same scale.}\label{S}
\end{figure}

\begin{figure}[!ht]
\plotone{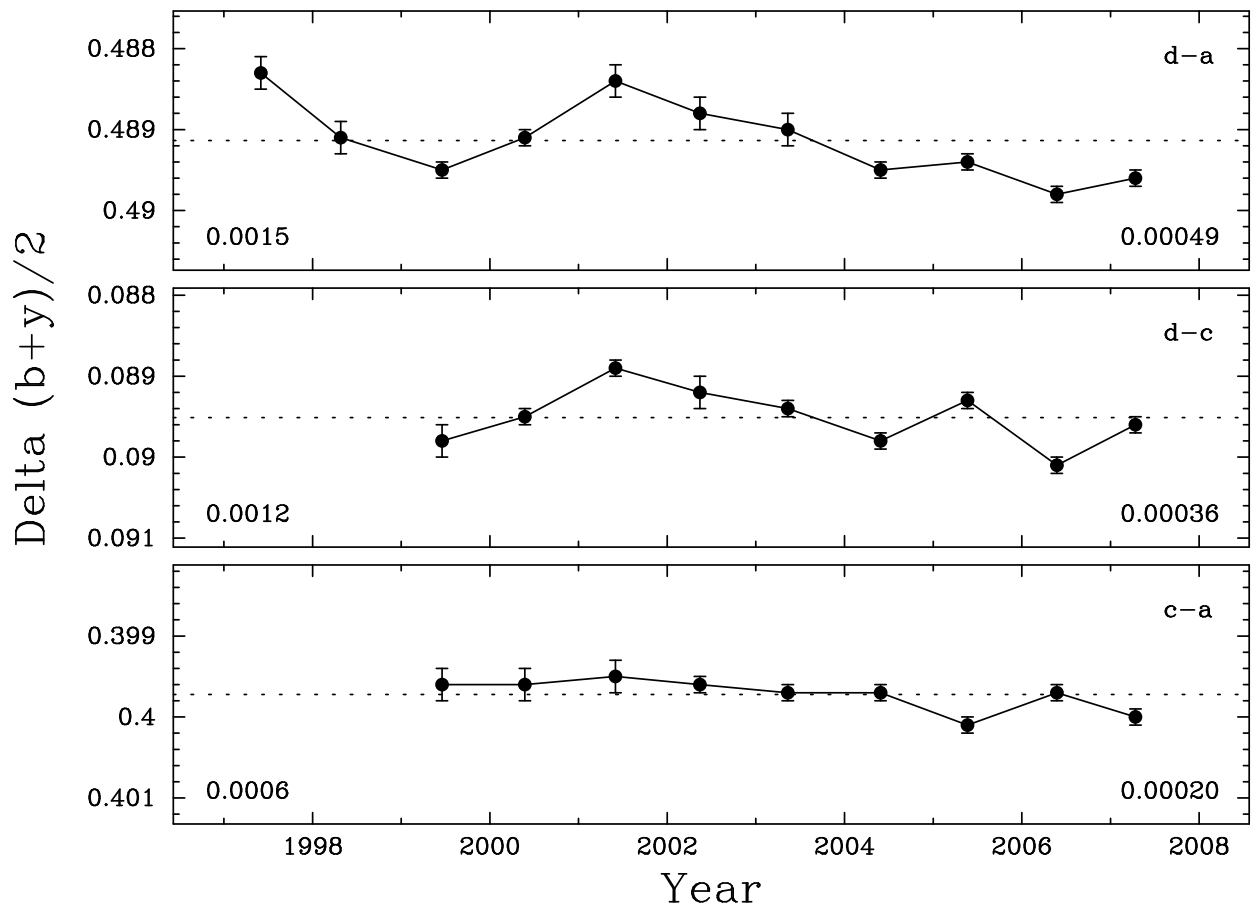}
\caption{\label{154345phot} Eleven years of Str\"omgren photometry for
  HD 154345 (star $d$) and two comparison stars ($a$ and $c$).  The
  plotted points represent yearly mean differential magnitudes between
  the stars indicated in the upper right of each panel.  The top two
  panels thus indicate the brightness variations of HD 154345 with
  respect to two photometric standards, and the bottom panel
  demonstrates the long-term stability of those standards.  The dotted
  line indicates the mean, and the number in the lower left corner
  indicates the total range of the plotted points.  The number in the
  lower right corner gives the standard deviation of the individual
  seasonal means about the mean.  A third comparison star ($b$) proved
  unsuitable for long-term photometric work, and comparison star ($c$)
  has only been observed since 1999.  Note that the period and phase
  of the photometric signal are the same as that of the chromospheric
  activity measurements in Fig.~\ref{S}.}
\end{figure}

We monitor all of our program stars for variations in chromospheric
activity, extracting Mount Wilson $S$-indices from our RV science
spectra \citep[H. Isaacson, {\it in preparation};][]{Wright04}.  We
have found that HD 154345 shows clear evidence of a stellar cycle.
Fig.~\ref{S} shows that the magnetic activity level of HD 154345
varies sinusoidally with a $\sim$ 9-year period.  Photometric
monitoring from Fairborn Observatory \citep{HenryG99} confirms the
presence of this activity cycle: Fig.~\ref{154345phot} shows a $\sim$
1 mmag photometric variation in phase with the chromospheric
emission\footnote{For G stars, visible luminosity varies on order
  0.1\% in phase with chromospheric emission such that the star is
  brightest during activity maximum \citep{Lockwood07}.}.

\citet{Deming87} observed an apparent drift in line centroids of CO
transitions at 2.3 $\mu$m in the integrated Solar spectrum over a
3-year span.  They associated these shifts with changes in the
activity level of the Sun over that period, and suggested that
activity cycles on Sun-like stars could thus mimic the RV signature of
a long-period exoplanet.  The apparent coincidence of the phase of HD
154345's magnetic cycle with that of the radial velocities in the same
sense at the shifts seen by \citet{Deming87} therefore demands that we
take a closer look at the effects of stellar activity cycles on radial
velocities measured in the optical.\footnote{A bisector analysis to
  distinguish line profile variations from true Doppler shifts is
  impractical for this system.  A Doppler shift can be described as a
  shift in the first moment of a line profile, distinguished from a
  profile change, which is a change in the higher moments.
  Measurements of line profile changes will therefore have inherently lower
  precision than measurements of Doppler shifts. We estimate that our
  sensitivity to these higher moments for our data in hand is only 25
  m/s, insufficient to distinguish them from 14 m/s Doppler shifts.}

The Mount Wilson H \& K activity survey \citep{Baliunas95b} identified
several stars of similar color to HD 154345 as exhibiting activity
variations consistent with cycles.  We have over 6 years of RV data at
Keck for a total of four of these stars\footnote{We have only 2.5
  years of data for a fifth star, HD 115616, too little to sample an
  entire activity cycle.  We also have 6 years of Lick velocities for
  a sixth star, HD 103095, which is a cycling subdwarf exhibiting a
  complex RV history with no long-term coherence.}: HD 26965, HD 3795,
HD 10476, and HD 185144, the latter three of which we have monitored
regularly at Keck for over 10 years.  HD 10476, HD 26965, and HD
185144 all have sufficiently strong cycles that we can confirm their
continued coherence since the end of the published Mount Wilson data
from our own activity measurements.

None of these four stars shows RV variations similar to HD 154345, or
any correlation of $S$-index with RV.  In fact, all of these stars
show r.m.s. RV variations of less\footnote{HD 3795, has a binary
  companion; we have therefore measured its r.m.s. RV variations after
  subtraction of a strong linear trend.} than 5 m \persec, and one, HD
185144, is among the most RV-stable stars in our entire sample.

\begin{figure}[!ht]
\plotone{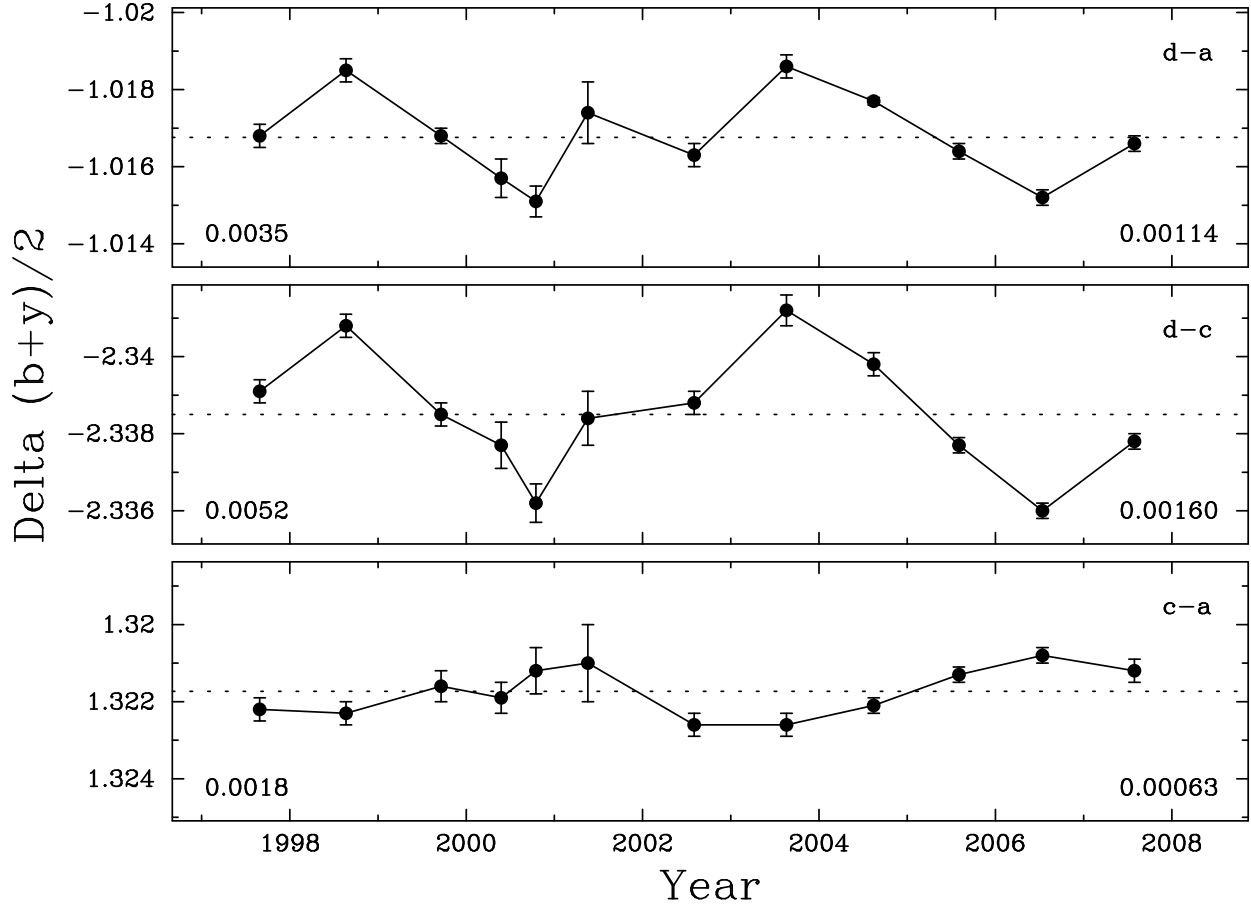}
\caption{As Fig~\ref{154345phot}, but for the RV stable star HD 185144
  (star $d$) and two different comparison stars $c$ and $a$.  Note
  that the photometric variations here are again consistent with the
  measurements in Fig.~\ref{S}, and show a higher amplitude than those
  of HD 154345.}\label{185144phot}
\end{figure}

We focus here on HD 185144 because its activity cycle is so clear and
it is one of the best-observed stars on our program.  Fig.~\ref{S}
shows that this star has a similar cycle period to that of HD 154345,
but with a higher mean activity level and larger
variations.\footnote{The phase and period we measure is consistent
  with that measured at Mount Wilson \citep[Fig.~1$f$ of][left column,
    fifth panel from top]{Baliunas95b}}.  Fig.~\ref{185144phot} shows
Fairborn Observatory photometry of HD 185144 which confirms the
existence, strength, period, and phase of this cycle.

Despite these clear activity variations, HD 185144 is one of the most
stable stars on our program.  We have monitored HD 185144 intensely at
Keck Observatory since 1997 and made more than 350 observations on
more than 60 nights over the past 10 years (since the activity and RV
measurements are taken from the same spectra, Fig.~\ref{S} also shows
the temporal coverage of our RV observations).  The r.m.s. scatter of
the RV variations over this entire period is less than 2.5 m \persec.

Based on these four stars, we conclude that long-term radial velocity
variations are generally not seen in the optical absorption lines of
late G stars undergoing magnetic activity cycles, even cycles as
strong as those in HD 185144.  We note that the Sun's 11-year activity
cycle has a period similar to that of Jupiter's orbit, and that the
Mount Wilson survey demonstrated that decadal activity cycles are a
common feature of old G stars \citep{Baliunas95b}.  We thus consider
the fact that HD 154345 exhibits an activity cycle with a period and
phase similar to that of its Jupiter twin to be an inevitable
coincidence.

\section{Discussion}

The orbital solution presented here is sensitive to the assumption
that the signal from any other planets in the system is not
significant.  The residuals to this fit show no significant
periodicities, consistent with there being no interior planets with
$\msini \gtrsim 0.3 \mjup$.  Our intense monitoring of this system at
1 m \persec\ precision since 2007 will provide increasingly stringent
limits on the existence of any interior giant planets.\footnote{The
  highest peaks in a periodogram of the residuals to this fit are near
  18, 40, and 45 days and have formal false alarm probabilities of a
  few percent.  The best two-planet fits at these periods involve
  planets on eccentric orbits with $\msini \sim 25 \mearth$.  While
  signals such as these are intriguing, they are not persuasive, and
  any conclusions drawn from them would be highly speculative.}

One of the difficulties of having only observed a single orbit is in
excluding the possibility of a detectable exterior planet.  For
instance, a Saturn analog in this system would contribute a
low-amplitude, long-term signal which might be absorbed in the
eccentricity term of the orbital solution above.  Long-term monitoring
of this system will eventually break this degeneracy.

HD~154345{\it b} will likely not be detected by other methods within
the next few years.  Its large orbital distance makes transits
extremely unlikely, but interior planets may be detectable if their
inclination is favorable and their orbital radius small.  Measurement
of the $> 250$ $\mu$as astrometric signature of HD~154345{\it b} over
its 9-year orbit would determine the orientation of the orbital plane
of this system and resolve the $\sin{i}$ ambiguity in the RV-derived
mass\footnote{For randomly oriented orbits, the median value of
  $1/\sin{i}$ is 1.15, and the true value for any given system will be
  within 20\% of that number, with 68\% confidence.}.  The planet is
separated from its parent star by a maximum of 0\farcs2 with a
prohibitive contrast of $\sim 21$ magnitudes (the age of the planet
means it has little intrinsic infrared luminosity and at 4.2 AU the
planet will reflect and reprocess very little light).

The brightness and proximity of HD 154345, however, will make this
system a natural target for future spaceborne efforts, such as SIM and
TPF.  The large, apparently empty region interior to HD~154345{\it b}
should be a prime target for these and other efforts to detect
terrestrial planets in the Habitable Zone around a Sun-like star.

\acknowledgments

We wish to recognize and acknowledge the very significant cultural
role and reverence that the summit of Mauna Kea has always had within
the indigenous Hawaiian community.  We are most fortunate to have the
opportunity to conduct observations from this mountain.

This research has made use of the SIMBAD database, operated at CDS,
Strasbourg, France, and of NASA's Astrophysics Data System
Bibliographic Services, and is made possible by the generous support
of NASA and the NSF, including grant AST-0307493.

\clearpage


\begin{deluxetable}{cc}
\tablecolumns{2} \tablecaption{Stellar Properties of
  HD~154345\label{star}} \tablehead{\colhead{Parameter} &
  \colhead{Value}} \startdata Spectral Type & G8 V \\
 {\it Hipparcos}
ID & 83389 \\
 RA & 17$^{\mbox{h}}$02$^{\mbox{m}}$36$.\!\!^{\mbox{
    s}}$404\\
 Dec. & +47$^\circ$04\arcmin 54\farcs77\\
 B-V & 0.73 \\
 V
& 6.76 \\
 Distance & 18.06 $\pm$ 0.18 pc\\
 M$_{\mbox{V}}$ & 5.48
\\
 \teff & 5468 $\pm$ 44 K\\
 $\log{g}$ $[$cm$\mbox{s}^2]$ & 4.537
$\pm$ 0.06 \\
 \feh & -0.105 $\pm$ 0.03 \\
 $v\sin{i}$ & 1.21 $\pm$ 0.5
km \persec\\
 Mass & 0.88 $\pm$ 0.09 \Msol \\
 Radius & 0.94 $\pm$ 0.03
R$_{\astrosun}$\\
 S & 0.18 \\
 \rphk & -4.91 \\
 \dmv & -0.21 mag
\enddata
\end{deluxetable}

\begin{deluxetable}{cc}
\tablecolumns{2} \tablecaption{Orbital Properties of HD~154345{\it
    b}\label{orbit}} \tablehead{\colhead{Parameter} & \colhead{Value}}
\startdata $P$ & 9.15 $\pm$ 0.26 yr\\
 $e$ & 0.044 $\pm$
0.046\tablenotemark{\dagger}\\
 $\omega$ & $68^\circ$
\tablenotemark{\dagger}\\
 $T_{\mbox{p}}$ (JD) & 2452830 $\pm$
330\\
 $K$ & 14.03 $\pm$ 0.75 m \persec\\
 \msini & 0.947 $\pm$ 0.090
\mjup\\
 $a$ & 4.19 $\pm$ 0.26 AU\\
 r.m.s. & 2.7 m
\persec\\
 $\chi^2_\nu$ & 0.89 \\
 N$_{\mbox{obs}}$ &
55\\
 N$_{\mbox{obs, binned}}$ & 41 \enddata \tablecomments{Errors in
  the orbital elements assume no other planets in the system and a
  uniform stellar jitter of 2.5 m \persec\ added after grouping the
  data in 3-day bins.}  \tablenotetext{\dagger}{Eccentricity
  consistent with zero.  See \citet{Butler06} for error explanation.
  Holding $e$ fixed at 0 yields a fit with nearly identical
  parameters, differing by no more than 1 m \persec\ from this one.}
\end{deluxetable}

\begin{deluxetable}{ccc}
\tablecolumns{3} \tablecaption{Radial Velocities for
  HD~154345\label{vels}} \tablehead{\colhead{Time} &
  \colhead{Velocity} & \colhead{Uncertainty} \\
 \colhead{JD-2440000} &
  \colhead{m \persec} & \colhead{m \persec} }

\startdata 10547.11003 & -14.5 &1.6\\
 10603.95584 & -12.7&1.4\\
 10956.01562 & -6.2 &1.6\\
 10982.96363 & -8.2 &1.4\\
 11013.86865 & -5.0 &1.5\\
 11311.06548 & -4.7 &1.6\\
 11368.78949 & -2.6 &1.6\\
 11441.71387 & 1.7 &1.3\\
 11705.91783 & 9.0 &1.6\\
 12003.07818 & 9.0 &2.2\\
 12098.91653 & 15.8 &1.6\\
 12128.79781 & 13.4 &1.6\\
 12333.17329 & 16.8 &1.7\\
 12487.86019 & 14.2 &1.5\\
 12776.98546 & 8.6 &1.7\\
 12806.95185 & -1.8 &1.6\\
 12833.80103 & 6.0 &1.4\\
 12848.77203 & 4.0 &1.5\\
 12897.77656 & 5.2 &1.6\\
 13072.04692 & -2.4 &1.6\\
 13074.07776 & -0.1 &1.6\\
 13077.12809 & -0.9 &1.5\\
 13153.94317 & -6.1 &1.7\\
 13179.99245 & -0.7 &1.6\\
 13195.81919 & -3.8 &1.5\\
 13428.16210 & -10.50 &0.81\\
 13547.91395 & -8.12 &0.81\\
 13604.83043 & -12.90 &0.76\\
 13777.15534 & -13.6 &1.5\\
 13807.07725 & -15.5 &1.5\\
 13931.95487 & -19.99 &0.66\\
 13932.91301 & -19.21 &0.65\\
 14248.02704 & -11.4 &1.5\\
 14248.99488 & -9.1 &1.4\\
 14249.94937 & -14.0 &1.4\\
 14252.03683 & -10.0 &1.1\\
 14255.93248 & -11.4 &1.1\\
 14277.86162 & -3.9 &1.6\\
 14278.90519 & -3.8 &1.4\\
 14279.94294 & -1.7 &1.5\\
 14285.90395 & -6.2 &1.5\\
 14294.89078 & -11.8 &1.4\\
 14304.87636 & -11.60 &0.88\\
 14305.88185 & -11.27 &0.92\\
 14306.87718 & -10.01 &0.88\\
 14307.92613 & -12.78 &0.90\\
 14308.90402 & -10.42 &0.87\\
 14309.88576 & -12.57 &0.86\\
 14310.87849 & -6.31 &0.79\\
 14311.86895 & -7.44 &0.80\\
 14312.86461 & -5.47 &0.80\\
 14313.86609 & -7.74 &0.92\\
 14314.89786 & -5.98 &0.87\\
 14318.93217 & -8.37 &0.78\\
 14335.81712 & -8.71 &0.91\\
 \enddata \tablecomments{The
  orbital fit in Table~\ref{orbit} assumes an additional 2.5 m
  \persec\ of stellar jitter.}
\end{deluxetable}

\end{document}